\documentclass{acm_proc_article-sp}

\usepackage{balance}

\begin{document}
\title{Improving Air Interface User Privacy in Mobile Telephony}

\author{Mohammed Shafiul Alam Khan \and Chris J Mitchell\\\\Information Security Group, Royal Holloway,
University of London\\Egham, Surrey TW20 0EX, United Kingdom\\\\
\email{shafiulalam@gmail.com, me@chrismitchell.net}}

\maketitle

\begin{abstract}

Although the security properties of 3G and 4G mobile networks have significantly improved by comparison with 2G (GSM), significant shortcomings remain with respect to user privacy. A number of possible modifications to 2G, 3G and 4G protocols have been proposed designed to provide greater user privacy; however, they all require significant modifications to existing deployed infrastructures, which are almost certainly impractical to achieve in practice. In this article we propose an approach which does not require any changes to the existing deployed network infrastructures or mobile devices, but offers improved user identity protection over the air interface. The proposed scheme makes use of multiple IMSIs for an individual USIM to offer a degree of pseudonymity for a user. The only changes required are to the operation of the authentication centre in the home network and to the USIM, and the scheme could be deployed immediately since it is completely transparent to the existing mobile telephony infrastructure. We present two different approaches to the use and management of multiple IMSIs. 

\end{abstract}

\category{}{Networks}{Network properties - network security - mobile and wireless security and network privacy and anonymity}


\terms{Security}


\section{Introduction}

While the first generation (1G) mobile telephony systems did not provide any security features, security has been an integral part of such systems since the second generation (2G).  For example, GSM, perhaps the best known 2G system, provides a range of security features, including authentication of the mobile to the network, data confidentiality across the air interface, and a degree of user pseudonymity through the use of temporary identities. Third and fourth generation (3G and 4G) systems, such as UMTS/3GPP and Long-Term Evolution (LTE), have enhanced these security features, notably by providing mutual authentication between network and phone, and integrity protection for signalling commands sent across the air interface.  However, user privacy protection has remained largely unchanged, relying in all cases on the use of temporary identities \cite{LTE-Security_Niemi-2012,UMTS-Security_Niemi_2003}, and it has long been known that the existing measures do not provide complete protection for the user identity \cite{ETSI_TS_121_133:2001,ETSI_TS_133_102:2013}. The discussion below applies equally to 2G, 3G and 4G systems, although we use 3G terminology throughout.  \\

In mobile telephony, each \emph{User Subscriber Identity Module (USIM)} has a unique \emph{International Mobile Subscriber Identity (IMSI)}\@. If the IMSI is sent in cleartext across the air interface, and the mapping to the phone user is known by an adversary, then a particular user could be tracked using the IMSI, since it is fixed for the lifetime of the USIM. To avoid this major privacy weakness, the visited network assigns the phone a \emph{Temporary Mobile Subscriber Identity (TMSI)}, which is used for addressing purposes across the air interface, and which changes frequently.  Of course, if an adversary could link a TMSI to the IMSI then this would represent a breach of user privacy. Unfortunately, as discussed in section \ref{User-Privacy} below, it is necessary to send the IMSI across the air interface in certain circumstances \cite{Privacy-in-Communication_Koien_2013}, e.g.\ when registering with the network after switching on a phone.  \\

This user privacy issue has been discussed extensively in the literature \cite{Arapinis12,LTE-Identity-Privacy_Choudhury_2012,deng_2009_novel,Privacy-Khan-2014,koien_2013_privacy,LTE-AN_Vintila_2011,xiehua_2011_security}, and many modifications to existing protocols have been proposed to avoid the problem \cite{Arapinis12,deng_2009_novel,juang_2007_efficient_AKA,xiehua_2011_security}. All these proposals involve making major modifications to the air interface protocol, which would require changes to the operation of all the serving networks as well as all the deployed phones. It seems likely that making the necessary major modifications to the operation of the air interface after deployment is essentially infeasible.  Many of the proposed schemes also involve the use of public key cryptography \cite{Arapinis12,deng_2009_novel,xiehua_2011_security}, which has a high computational cost, although there do exist schemes which only use symmetric cryptography \cite{juang_2007_efficient_AKA}. It would therefore be extremely valuable if a scheme offering greater user privacy could be devised which did not involve making significant changes to the existing mobile telecommunications infrastructures, and had minimal computational cost. This motivates the work described in this paper. \\

The remainder of the paper is structured as follows. In
section \ref{BG} the key terminology and features related to mobile telephony systems are briefly reviewed. Privacy terminology and the threats to mobile user privacy in existing networks are then summarised in section \ref{User-Privacy}. In section \ref{PTM} our threat model is presented. Section \ref{Proposal} outlines a novel approach to improving air interface user privacy in mobile telephony using multiple IMSIs. Sections \ref{Proposal-PM-IMSI} and \ref{Proposal-MM-IMSI} provide descriptions of two proposed approaches to the use and management of multiple IMSIs in a USIM. An analysis of the proposed approaches is presented in section \ref{analysis}. Section \ref{RW} provides a brief discussion of related work. Finally, conclusions are drawn and possible directions for future work are considered in section \ref{Conclusion}.

\section{Background} \label{BG}

\subsection{Mobile Systems} \label{MS}

In 3G systems a complete mobile phone is referred to as a \emph{User Equipment (UE)\@}, where the term encapsulates not only the \emph{mobile equipment (ME)\@}, i.e.\ the phone, but also the USIM within it \cite{3GPP_TR_21_905}, where the USIM takes the form of a cut-down smart card. The USIM embodies the relationship between the human user and the issuing \emph{home network}, including the IMSI, the telephone number of the UE, and other user (subscriber) data.  In particular the USIM holds a secret key shared with the issuing network, which forms the basis for all the air interface security features. \\

The USIM data storage capabilities are specified in section 10.1 of 3GPP TS 121 111 \cite{ETSI_TS_121_111:2008}. Information held within the USIM is stored in files, which are divided into the following categories:  \emph{application dedicated files (ADFs)}, \emph{dedicated files (DFs)} and \emph{elementary files (EFs)}, where EFs are further sub-divided into transparent EFs, linear fixed EFs and cyclic EFs depending on their storage requirements \cite{3GPP_TS_31_101}. Files are composed of a header, which is internally managed by the USIM, and, optionally, a body part. The information in the header is related to the structure and attributes of the file, and may be retrieved from the USIM by sending it specific commands. The body part contains the data of the file. Most of the subscriber information is stored in EFs, which are the files we focus on in this paper. \\

An IMSI is a fifteen numerical digit number. Of the fifteen digits, the first three form the mobile country code (MCC). The next two or three digits identify the network operator, and are known as the mobile network code (MNC). The length of the MNC, i.e.\ whether it contains two or three digits, is a national matter. The remaining nine or ten digits, known as the mobile subscriber identification number (MSIN), are administered by the relevant operator in accordance with the national policy \cite{3GPP_TS_23_003:2003,ITU_T_E_212}. IMSIs therefore have geographical significance, and their use is typically managed by the network operator in blocks. The combination of the MCC and the MNC can be used to uniquely identify the home network of the IMSI. The MSIN is used by the operator to identify the subscriber for billing and other operational purposes. Therefore, each IMSI uniquely identifies the mobile user, as well as the user's home network and home country. The IMSI is stored in the USIM and is normally fixed. The elementary file \emph{\(EF_{IMSI}\)} contains the value of the IMSI.   

\subsection{Proactive UICC}

\emph{Proactive UICC} is a service operating across the USIM-ME interface that provides a mechanism for a USIM to initiate an action to be taken by the ME\@. It forms part of the USIM application toolkit. The 2G predecessor of the USIM, known as the SIM, supports a similar feature, known as \emph{proactive SIM}, part of the SIM application toolkit. As described in the specification \cite{ETSI_TS_131_111}, proactive UICC extends proactive SIM. \\

ETSI TS 102 221 \cite{ETSI_TS_102_221} specifies that the ME must communicate with the USIM using either the T=0 or T=1 protocol, specified in ISO/IEC 7816-3 \cite{ISO_IEC_7816-3}. In both cases the ME is always the \emph{master} and thus initiates commands to the USIM; as a result there is no mechanism for the USIM to initiate communications with the ME\@. This limits the possibility of introducing new USIM features requiring the support of the ME, as the ME needs to know in advance what actions it should take. The proactive UICC service provides a mechanism that allows the USIM to indicate to the ME, using a response to a ME-issued command, that it has some information to send. The USIM achieves this by including a special status byte in the response application protocol data unit. The ME is then required to issue the \emph{FETCH} command to find out what the information is \cite{ETSI_TS_102_223}. To avoid cross-phase compatibility problems, this service is only permitted to be used between a proactive UICC and a ME that supports the proactive UICC feature. The fact that an ME supports proactive UICC is revealed when it sends a \emph{TERMINAL PROFILE} command during UICC initialization.  \\

The USIM can make a variety of requests using the proactive UICC service. Examples include: requesting the ME to display USIM-provided text, notifying the ME of changes to EF(s), and providing local information from the ME to the USIM \cite{ETSI_TS_102_223}. 
The commands of interest in this paper are \emph{REFRESH} and \emph{DISPLAY TEXT}. The \emph{REFRESH} command requests the ME to carry out an initialisation procedure, or advises the ME that the contents of EF(s) have been changed. The command also makes it possible to restart the session by performing a reset. The \emph{DISPLAY TEXT} command allows the UICC to request the ME to display text or an icon replacing anything else on its screen \cite{ETSI_TS_102_223}.

\subsection{The AKA Protocol}  \label{AKA}

At the core of mobile telephony air interface security is a mutual authentication and authenticated key establishment protocol known as \emph{AKA (Authentication and Key Agreement)}. This is regularly performed between the visited network and the UE\@. The involved parties in AKA are the home network (the network that issued the USIM), the serving network and the UE\@. The \emph{authentication center (AuC)} of the home network generates authentication vectors (authentication data used by the serving network in AKA) and sends them to the serving network. \\

The AKA protocol starts with the serving network sending an \emph{user
authentication request} to the UE\@. The UE checks the validity of
this request (thereby authenticating the network), and then sends a
\emph{user authentication response}. The serving network checks this
response to authenticate the UE\@.  As a result, if successful, the UE\@ and the network have authenticated each other, and at the same time they
establish two shared secret keys.\\

In order to participate in the protocol, the UE, in fact the USIM installed inside the UE, must possess two values:
\begin{itemize}
\item a long term secret key $K$, known only to the USIM and to the USIM's `home network', and
\item a sequence number \emph{SQN}, maintained by both the
    USIM and the home network.
\end{itemize}

The key $K$ never leaves the USIM, and the values of $K$ and
\emph{SQN} are protected by the USIM's physical security
features.  \\

The 48-bit sequence number \emph{SQN} enables the UE
to verify the `freshness' of the user authentication request.
More specifically, the request message contains two values:
\emph{RAND} and \emph{AUTN}, where \emph{RAND} is a 128-bit
random number generated by the home network, and the 128-bit
\emph{AUTN} consists of the concatenation of three values:
\emph{SQN}$\oplus$\emph{AK} (48 bits), \emph{AMF} (16 bits),
and \emph{MAC} (64 bits). The MAC is a Message Authentication Code (or \emph{tag})
computed as a function of \emph{RAND}, \emph{SQN}, \emph{AMF},
and the long term secret key $K$, using a MAC algorithm known
as $f$1. The value \emph{AK} is computed as a function of
$K$ and \emph{RAND}, using a cipher mask generating function known as $f$5. 
The \emph{AK} functions as a means of
encrypting \emph{SQN}; this is necessary since, if sent in
cleartext, the \emph{SQN} value would potentially compromise
user identity confidentiality, given that the value of
\emph{SQN} is USIM-specific.\\

On receipt of these two values, the USIM uses the received
\emph{RAND}, along with its stored value of $K$, to regenerate the
value of \emph{AK}, which it can then use to recover \emph{SQN}\@.
It next uses its stored key $K$, together with the received values
of \emph{RAND}, \emph{SQN}, and \emph{AMF}, in function $f$1 to regenerate \emph{MAC}; if the newly computed value agrees with the value
received in \emph{AUTN} then the first stage of authentication has
succeeded. The USIM next checks that \emph{SQN} is a `new' value; if
so it updates its stored \emph{SQN} value and the network has been
authenticated. \\

If authentication succeeds, the USIM computes another message
authentication code, called \emph{RES}, from $K$ and \emph{RAND}
using a distinct MAC function $f$2, and sends it to the network as
part of the user authentication response. If this \emph{RES} agrees
with the value expected by the network then the UE is deemed
authenticated.

\section{User Privacy} \label{User-Privacy}

We start our consideration of user privacy by introducing the privacy-related terminology of relevance to this paper. \emph{Anonymity} of a subject means that the subject is not identifiable within a set of subjects known as the anonymity set \cite{Privacy-Term_Hansen_2012}. \emph{Unlinkability} of two or more items of interest (e.g.\ subjects, messages, actions) means that within the system an interested party cannot distinguish whether or not they are related \cite{Privacy-Term_Hansen_2012}. \emph{Untraceability} of an item of interest means that an interested party cannot determine whether or not it exists or is present. User anonymity, unlinkability and untraceability are clearly all desirable properties from a user privacy perspective.  \\

In a mobile telephony context, a user identity can be the mobile number or the IMSI of a USIM, or the international mobile station equipment identity (IMEI) of an ME\@. Of these various identities, the IMSI is used to identify the subscriber for authentication, access provision and accounting purposes; limiting the degree to which its use compromises user privacy is the main focus of this paper. When a subscriber is roaming, i.e.\ accessing service from a network other than its home network, the IMSI is sent from the UE via the visited network to the home network. Since the IMSI is a permanent user identity, the air interface protocols are designed to minimise the number of circumstances in which it is sent across the air interface. \\

Clearly, providing identity confidentiality in mobile telephony requires that the permanent user identity cannot be intercepted when sent across the radio access link. A similar requirement applies to the provision of user location confidentiality and user untraceability, which, like identity confidentiality, are identified as desirable features in the 3GPP security architecture  \cite{3GPP_TS_33_102}. A level of identity confidentiality is provided by use of the TMSI instead of the IMSI. However, as noted above, there are circumstances where a UE needs to send its IMSI across the air interface in cleartext. One such case is when a UE is switched on and wishes to connect to a new network, and hence will not have an assigned TMSI. Another case is where the serving network is unable to identify the IMSI from the TMSI \cite{ETSI_TS_133_102:2013}. An active adversary can intentionally simulate one of these scenarios to force a UE to transfer its IMSI in cleartext. Moreover, several further scenarios have been identified  \cite{Arapinis12,ETSI_TS_121_133:2001,LTE-AN_Vintila_2011} in which use of the TMSI fails to protect user identity confidentiality. In the remainder of this paper, we propose a novel approach to improving user privacy in the air interface, which requires no modifications to the existing deployed mobile telecommunications infrastructures, including the phones. Before doing so we present our threat model.

\section{Threat Model} \label{PTM}

The schemes we propose in this paper are designed to address real-life threats to user privacy in 3G networks. In particular we have already observed that there are circumstances in which an opponent can cause a UE to send its IMSI across the network in plaintext. This is the threat we aim to mitigate, by reducing the impact of IMSI compromise. That is, although the possibility of IMSI compromise remains unchanged, the schemes we propose make the IMSI a short term identity, and hence we prevent the compromise of a long-term user identity. In doing so we must also ensure that two different IMSIs for the same UE are not linkable, at least via the network protocol. This issue is examined further in section \ref{user-privacy}. \\

In designing our schemes we make the underlying assumption that the AKA protocol is sound, and provides mutually authenticated key establishment. We also implicitly assume that the USIM and the network have not been compromised by other means. Of course, if these assumptions are false, then very serious threats exist to both user privacy and security. Since we assume that the AKA protocol is secure, and no changes are made to its operation, we do not need to re-examine its security for the schemes we discuss below. The main risk introduced by use of the multiple-IMSI schemes we propose is the possibility of loss of IMSI synchronisation between UE and home network, and this issue is addressed in section \ref{IMSI-synch}. \\

It is important to note that two of the three schemes we describe rely on using the \emph{RAND} sent as part of the AKA protocol as a means of signalling from the home network to the USIM. We observe that, from our assumption regarding the security of AKA, we can assume that this provides an authenticated channel with replay detection. This is fundamental to the schemes presented in sections \ref{HN_Init} and \ref{Proposal-MM-IMSI}.

\section{A Pseudonymity Approach} \label{Proposal}

The capabilities of the USIM continue to evolve, allowing the possible storage of multiple IMSIs \cite{patent_identity}. Whilst other authors have proposed to use this capability to allow the storage of multiple virtual USIMs on a single device, as discussed in section \ref{RW}, we consider here the possible use of multiple IMSIs for a single account to provide a form of pseudonymity on the air interface, even when it is necessary to send the IMSI in cleartext. The use of multiple IMSIs is described here using 3G terminology, but a precisely analogous approach would apply equally to both GSM and LTE systems. However, while all the techniques for IMSI distribution specified in sections \ref{Proposal-PM-IMSI} and \ref{Proposal-MM-IMSI} would also work for LTE, only the scheme described in section \ref{USIM_Init} would work for GSM, since the other two schemes rely on UE authentication of the network which is not provided in GSM\@.\\ 

At present, a USIM holds one IMSI with other relevant parameters specific to the subscription and the network. We propose that a USIM and the home network support the use of varying IMSIs for a single user account, in such a way that no modification is required to the operation of any intermediate entities, notably the visited (serving) network and the ME itself. This allows the provision of a more robust form of pseudonymity without making any changes to the air interface protocol. In this section we consider in greater detail how the change of IMSI can be made. \\

A number of issues need to be addressed in order to make use of multiple IMSIs, as follows.
\begin{itemize}

\item \emph{Transferring IMSIs.} Clearly, before a USIM switches to a new IMSI, this value must be present both in the USIM and in the database of the home network. Moreover, new IMSIs must always be chosen by the home network to avoid the possibility of a single IMSI being assigned to two different USIMs. This requires a direct means of communication between the home network and the USIM (which must be transparent to the serving network and the ME, since our objective is to enable changing of an IMSI without making any changes to existing deployed equipments). In sections \ref{Proposal-PM-IMSI} and \ref{Proposal-MM-IMSI} we describe in detail two different strategies for transferring IMSIs from the home network to a USIM.

\item \emph{Initiating an IMSI change.} Clearly the IMSI needs to be changed in such a way that both the home network and the USIM know at all times which IMSI is being used, and the home network always knows the correspondence between the IMSI being used by the USIM and the user account. An IMSI change can be triggered either by the USIM or by the home network, as we describe below. However, use of a new IMSI is always implemented by the USIM, since it is the appearance of a mobile device in a network using a particular IMSI which causes a request to be sent by the serving network for authentication information for use in the AKA protocol. That is, when the ME sends an IMSI to the serving network, it is forwarded to the home network.  Once the home network sees the `new' IMSI it knows that an IMSI change has occurred and can act accordingly.

This requires that the home network knows that both the previously used IMSI and the `new' IMSI belong to the same account. This will require some minor changes to the operation of the home network's account database, i.e.\ to allow more than one IMSI to point to a single account. However, this does not seem likely be a major problem in practice.

\item \emph{Triggering an IMSI changes.} Whether the USIM or the home network is responsible for initiating a change of IMSI, logic needs to be implemented to cause such a change to take place. Regardless of whether the USIM or the home network makes the decision, logic needs to be in place in the USIM either to make the decision or to receive the instruction to make the change from the home network; for convenience we refer to this logic as an application, although this is not intended to constrain how it is implemented. The decision-making logic could take account of external factors, including, for example, the elapsed time or the number of AKA interactions since the last change; indeed, if the ME included an appropriate user-facing application, then it might also be possible to allow user-initiated changes. Of course, if the home network is responsible for triggering the change of IMSI, then it needs a means of communicating its decision to the USIM that is transparent to the existing infrastructure, including the serving network and the ME. This issue is addressed in sections \ref{Proposal-PM-IMSI} and \ref{Proposal-MM-IMSI}.

\item \emph{Rate of change of IMSI.} The rate of change of IMSI will clearly be decided by the network which issues the USIM (and equips it with the IMSI-changing application). We observe in this context that section 4.2.2 of ETSI TS 131 102 \cite{ETSI_TS_131_102} recommends that IMSI updates should not occur frequently. The rate of change of an IMSI could be determined by the customer contract with the issuing network; for example, a USIM which changes its IMSI frequently could cost more than a fixed-IMSI USIM (or one that only changes its IMSI occasionally), and could be marketed as part of a special `high-privacy' service.

\item \emph{Implementing an IMSI change.} A mechanism will be required for the USIM to indicate to the ME that the IMSI has changed.  We propose that this should be achieved by the following steps.
\begin{enumerate}

\item As noted in section \ref{MS}, the IMSI is contained in the elementary file \emph{\(EF_{IMSI}\)}. When the USIM wishes to change the IMSI, it first updates this file accordingly.
\item At the first opportunity, the USIM uses the proactive UICC status byte to indicate to the ME that it wishes to issue a command.
\item When the ME responds with a \emph{FETCH} command, the USIM sends a \emph{REFRESH} or a \emph{DISPLAY TEXT} command to the ME, depending on the capabilities of the ME.
\item The \emph{REFRESH} command causes the ME\@ to read the \emph{\(EF_{IMSI}\)} file, allowing it to discover the new IMSI. Alternatively, the \emph{DISPLAY TEXT} command causes the ME to display text instructing the user to restart the ME.
\item On the next occasion the ME is required to send its IMSI to the serving network, it will send the new value.

\end{enumerate}

\end{itemize}

As noted above, the use of multiple IMSIs requires a direct and transparent means of communication between the home network and the USIM. We observe that the \emph{Unstructured Supplementary Service Data (USSD)} protocol appears at first sight to be a possible channel for such communications. However, the protocol end points are the home network and the ME, rather than the USIM. As such, it could only be used for our purposes if the ME was aware of the multiple IMSI scheme, i.e.\ the ME would need to be modified --- contradicting our design objectives. As a result we do not consider the use of USSD further here, although we note that it might be possible to deploy a smart phone application which could provide the necessary additional phone functionality, a possible avenue for future research.

\section{Predefined Multiple IMSIs} \label{Proposal-PM-IMSI}

Our first proposed means of deploying multiple IMSIs involves an issued USIM being pre-equipped with a number of IMSIs. All these IMSIs are associated with a single account in the home network's account database.  Initially, one of these IMSIs is stored in the file \emph{\(EF_{IMSI}\)}\@. We propose below two different approaches to initiating a change of IMSI in this case.

\subsection{USIM-initiated IMSI change}  \label{USIM_Init}

This is the simpler of the two approaches. We suppose that the USIM is equipped with an application that decides when to initiate an IMSI change. The application could make this decision on the basis of use of the USIM by the ME, e.g.\ the number of times the AKA protocol is performed. \\

The new IMSI will clearly need to be selected from the pre-programmed list. How the list is used would be a matter for the issuing network. For example, the IMSIs could be used in cyclic order, or they could be used at random (or, more probably, pseudo-randomly). The USIM changes the IMSI to the `new' IMSI using the procedure described in section \ref{Proposal}.

\subsection{Network-initiated IMSI change}  \label{HN_Init}  

In this case, the home network decides when to trigger an IMSI change. The home network will have a richer set of information available to use to decide when to make an IMSI change than the USIM\@. For example, the home network could decide to change an IMSI whenever the USIM changes serving network, or after a fixed number of calls. \\

\begin{figure*}
\centering
\epsfig{file=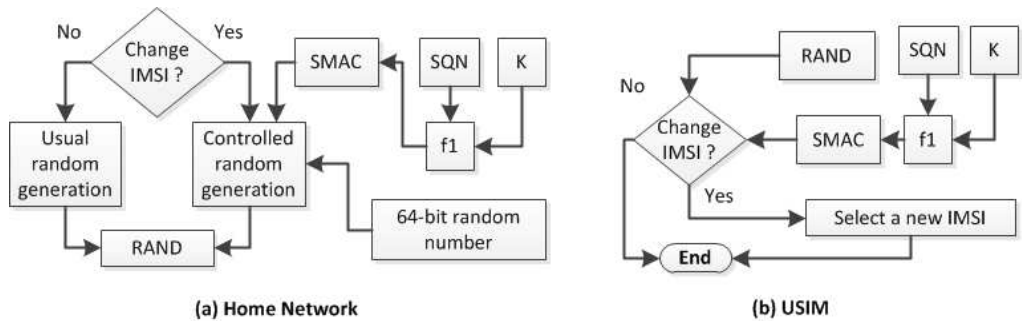}
\caption{IMSI change procedure for predefined multiple IMSIs}
\label{fig:P-IMSI}
\end{figure*}

As discussed in section \ref{Proposal}, when the home network decides to trigger an IMSI change, it must, by some means, send an instruction to the USIM\@. We propose to use the AKA protocol as the communications channel for this instruction. More specifically, we propose using the challenge value (\emph{RAND}) of AKA to carry the signal. The IMSI change procedure operates as follows (see also Figure~\ref{fig:P-IMSI}).

\begin{enumerate}
  
\item When the logic in the home network decide that an IMSI change is necessary, a flag is set for the appropriate user account in the AuC database of the home network. 
\item Whenever the AuC is required to generate authentication vectors for use in the AKA protocol, it checks this flag to see if an IMSI change signal is to be embedded in the \emph{RAND} value. If so it resets the flag and follows the following steps (as shown in Figure~\ref{fig:P-IMSI}(a)).

\begin{enumerate}
\item The AuC uses the existing MAC function $f$1\footnote{For cryptographic cleanliness it should be verified that the data string input for this additional use of $f$1 can never be the same as the data string input to $f$1 for its other uses; alternatively, a slight variant of $f$1 could be employed here.} to generate a 64-bit \emph{MAC} on the subscriber's current sequence number \emph{SQN}\@ using the subscriber's long term cryptographic key $K$\@. We refer to this as the sequence-MAC or \emph{SMAC}\@. 
\item The AuC generates a 64-bit random number \emph{R} using the same process as normally used to generate 128-bit \emph{RAND} values.
\item The AuC sets \emph{RAND} to be the concatenation of the \emph{R} and \emph{SMAC}.
\end{enumerate}

If an IMSI change signal is not required, the AuC generates \emph{RAND} in the normal way.

\item The AuC follows the standard steps to generate the authentication vector from the \emph{RAND} value, and sends the vector (including \emph{RAND}) to the serving network.
\end{enumerate}

Whenever the USIM receives an authentication request, it follows the usual AKA steps. If the AKA procedure completes successfully, the USIM checks the \emph{RAND} in the following way (as shown in Figure~\ref{fig:P-IMSI}(b)). 
 
\begin{enumerate}

\item The USIM uses the received \emph{SQN} and its stored long term key $K$ to regenerate \emph{SMAC}\@.
\item It compares the computed \emph{SMAC} with the appropriate part of \emph{RAND}\@.
\item If they do not agree then the USIM terminates the checking process. However, if they agree then the USIM performs the next step. 
\item The USIM selects a `new' IMSI value from the stored list, and changes the IMSI accordingly using the procedure described in section \ref{Proposal}.

\end{enumerate}

We next consider how the IMSI change process will work in practice. There are two cases to consider. If the home network is the same as the serving network then the home network could potentially force an instance of the AKA protocol to occur at will, i.e.\ making the IMSI change happen almost immediately. However, if the serving network is distinct from the home network, then the home network will only be able to send new authentication vectors when they are requested by the serving network. Moreover, the serving network may wait for a while before using the supplied authentication vector in the AKA protocol.  That is, in this case there may be a significant delay between the decision being made to change an IMSI and the signal being sent to the USIM\@. In either case the phone may actually be switched off or temporarily out of range of a base station, in which case there will inevitably be some delay. However, regardless of the length of the delay in the signal reaching the USIM (or even if the signal never reaches the USIM) there is no danger of loss of IMSI synchronisation between the USIM and the home network, since the home network will always retain the complete list of IMSIs allocated to the particular USIM\@. \\

We observe that there is always the chance that a randomly chosen \emph{RAND} will contain the `correct' \emph{SMAC}\@, leading to an unscheduled IMSI change by the USIM. However, the probability of this occurring is \(2^{-64}\), which is vanishingly small. In any case, the occurrence of such event would not have an adverse impact, since the home network would always be aware of the link between the new IMSI and the particular USIM\@.   \\

Finally, an active interceptor could introduce its own \emph{RAND} into the channel in an attempt to force an IMSI change. However, given that $K$ is not compromised and $f$1 has the properties required of a good MAC function, then no strategy better than generating a random \emph{RAND} will be available. Replays of old \emph{RAND} values will be detected and rejected as a normal part of the AKA protocol (at least for 3G and 4G networks, which enable the USIM to check the freshness of an authentication request). Finally, assuming the \emph{SMAC} value is indistinguishable from a random value, a standard assumption for MAC functions, then an eavesdropper will be unable to determine when an IMSI change is being requested.

\section{Modifiable Multiple IMSIs} \label{Proposal-MM-IMSI}

The second proposed means of deploying multiple IMSIs involves distributing new IMSI values from the home network to the USIM after its initial deployment, where the home network will choose each new IMSI from its pool of unused values. Such an approach clearly requires a means of communicating from the home network directly to the USIM, and, analogously to the scheme proposed in section \ref{HN_Init}, we describe how the AKA protocol, and specifically the \emph{RAND} value, can be used for this purpose. \\ 

Before describing the details of the IMSI transfer procedure, we describe some relatively minor changes which are required to the operation of the home network in order to support the scheme.

\begin{itemize}
\item The home network must maintain a pool of unused IMSIs, enabling the AuC to dynamically assign a new IMSI to an existing subscriber.  

\item For each subscriber account in its database, the home network must maintain an \emph{IMSI-change} flag indicating whether an IMSI change is under way. The database must also hold up to two IMSIs for each subscriber; it will always hold the current IMSI (with status \emph{allocated}) and, if the \emph{IMSI-change} flag is set, it may also hold the new IMSI (with status \emph{in transit}), where the possible status values for an IMSI are discussed below. If use of the new IMSI is observed then IMSI status changes are triggered (see below).

\item The home network must manage the use of IMSIs so that no IMSI is assigned to more than one subscriber at any one time. This can be achieved by maintaining the status of each IMSI as one of \emph{allocated}, \emph{free}, or \emph{in transit}. The set of IMSIs with status \emph{free} corresponds to the pool of available IMSIs referred to above. The status of an IMSI can be updated in the following ways.

\begin{itemize}
	\item When the home network selects an available IMSI from the pool to allocate to a USIM, the home network changes the status from \emph{free} to \emph{in transit}.
	\item When the home network receives implicit acknowledgement (in the form of a request for authentication vectors for that IMSI from a network) of a successful IMSI change, the home network changes the status of the IMSI from \emph{in transit} to \emph{allocated}, and the status of the previously used IMSI  for that subscriber from \emph{allocated} to \emph{free}. In addition, the current IMSI for the subscriber will be set equal to the new IMSI, the new IMSI will be set to null, and the \emph{IMSI-change} flag will be reset.
	\item A third case also needs to be considered, that is when an IMSI change instruction never reaches the USIM. If this case is not addressed then future IMSI changes for that USIM will be blocked. On the other hand, making a decision to abandon an IMSI change could be disastrous, i.e.\ if a USIM makes an IMSI change after the home network has terminated this change (and changed the status of the `new' IMSI back to \emph{free}), then the USIM could be rendered unusable. As a result we propose never to abandon an IMSI change, and instead to resend the new IMSI as many times as necessary until the change is accepted by the USIM. How this works should be clear from the description below.

\end{itemize}

\item If the home network is required to do so by its regulatory environment, e.g.\ to support lawful interception, it can maintain a log of all the IMSIs assigned to a particular subscriber for however long is required. It is also likely to be necessary to retain this information for a period to enable processing of billing records received from visited networks.

\end{itemize}

\begin{figure*}
\centering
\epsfig{file=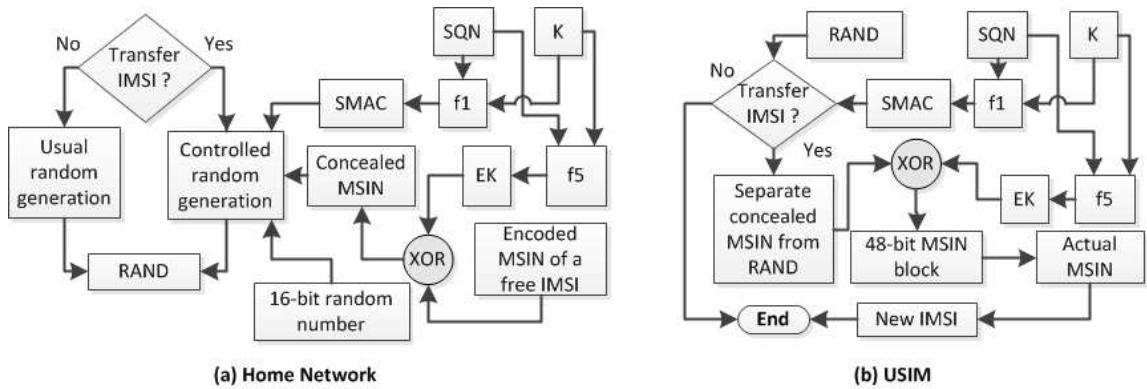}
\caption{IMSI change procedure for modifiable multiple IMSIs}
\label{fig:M-IMSI}
\end{figure*}

The details of the IMSI transfer procedure are as follows (see also Figure~\ref{fig:M-IMSI}).

\begin{enumerate}
  
\item When the logic in the home network decides that an IMSI transfer is necessary for a particular subscriber, it must set the \emph{IMSI-change} flag for that subscriber. Observe that if an IMSI change is already under way then the flag will already be set; in this case the flag is left as it is.

\item Whenever the AuC is required to generate authentication vectors for use in the AKA protocol, it checks this flag to see if an IMSI transfer signal and a new IMSI are to be embedded in the \emph{RAND} value. If so it follows the following steps (as shown in Figure~\ref{fig:M-IMSI}(a)). Note that this means that, once an IMSI change has been initiated, the new IMSI will be embedded in all \emph{RAND} values until evidence of the successful changeover by the USIM has been observed.

\begin{enumerate}
	\item The AuC uses the MAC function $f$1 to generate a 64-bit \emph{MAC} on the subscriber's current sequence number \emph{SQN} using the subscriber's long term cryptographic key $K$. The generated MAC is referred as \emph{SMAC}\@.
	\item The AuC generates a 48-bit encryption key \emph{EK} using the key generation function $f$5. The function takes \emph{SQN} as the data input and $K$ as the key input. Note that observations regarding cryptographic cleanliness and the use here of functions $f$1 and $f$5, analogous to those given in section \ref{HN_Init} step 2(a), apply here.
 	\item If the new IMSI field in the home network database entry for this subscriber is non-null then a new IMSI has already been assigned, and it is not necessary to choose another new value. Otherwise a new IMSI is selected from the pool of unused IMSIs; the status of this IMSI is changed from \emph{free} to \emph{in transit} and the new IMSI field in the database is given the chosen value. We assume that the MCC and MNC of the IMSI are known to the USIM (since they are fixed for this network operator) and hence only the 9- or 10-digit MSIN needs to be sent embedded in \emph{RAND}. The MSIN is encoded as a 36- or 40-bit value using binary coded decimal, the `standard' way of encoding IMSIs, and the result is padded to 48 bits by an agreed padding scheme.
 	\item The 48-bit MSIN block is XORed with the encryption key \emph{EK}\@, and we refer to the result as the concealed MSIN.    	
	\item The AuC generates a 16-bit random number \emph{R} using the same process as normally used to generate 128-bit \emph{RAND} values.
	\item The AuC sets \emph{RAND} to be the concatenation of the concealed MSIN, \emph{R} and \emph{SMAC}.
\end{enumerate}

If an IMSI transfer is not required, the AuC generates \emph{RAND} in the normal way.

\item The AuC follows the standard steps to generate the authentication vector from the \emph{RAND} value, and sends the vector (including \emph{RAND}) to the serving network.

\end{enumerate}

On receipt of an authentication request, the USIM proceeds using the standard AKA procedure. After successful completion of the AKA protocol, the USIM checks whether the challenge value contains an embedded IMSI in the following way (as shown in Figure~\ref{fig:M-IMSI}(b)). 

\begin{enumerate}

\item The USIM uses the received \emph{SQN} and its stored long term key $K$ to regenerate \emph{SMAC}\@.
\item It compares the computed \emph{SMAC} with the appropriate part of \emph{RAND}\@.
\item If they do not agree then the USIM terminates the checking process. However, if they agree then the USIM performs the following steps.

\begin{enumerate}
	\item The USIM retrieves the concealed MSIN from \emph{RAND}\@. 	
	\item The USIM regenerates the encryption key \emph{EK} using $f$5 with the value of \emph{SQN} retrieved during the AKA processing and its long-term stored key $K$ as inputs.
	\item The \emph{EK} is XORed with the concealed MSIN to recover the cleartext encoded MSIN.
	\item The USIM generates the new IMSI by prefixing the decoded MSIN with the MCC and MNC.
	\item The USIM checks whether the new IMSI is the same as the value it is using already; this is essential since it may receive the change instruction more than once. If they are the same it does nothing. If they are different it updates its IMSI using the procedure described in section \ref{Proposal}.
	
\end{enumerate} 

\end{enumerate}

To reduce signalling costs, it appears to be standard practice for the AuC to generate a small set of authentication vectors for provision to a serving network. If the procedure specified above is followed to generate this set of vectors, and an IMSI change is scheduled for the subscriber, then all the \emph{RAND} values in the set will contain an embedded concealed MSIN. Whilst this will cause minimal additional overhead for the USIM, since \emph{RAND} values are always checked for an embedded \emph{SMAC} value, it will have the benefit of maximising the chance that the IMSI change will be performed by the USIM. \\

As discussed in section \ref{HN_Init}, there may be a significant delay in the IMSI change signal embedded in the \emph{RAND} reaching the USIM. However, this will not affect IMSI synchronisation between the home network and the USIM since the home network will not update the current IMSI entry in the subscriber database until it receives a request for authentication vectors from a visited network using this new IMSI. As discussed above, once a new IMSI has been assigned to a subscriber (with the \emph{in transit} status), every \emph{RAND} generated for that USIM will contain the embedded IMSI value until the success of the change has been observed. \\

Finally, as discussed in section \ref{HN_Init}, there is the chance that a randomly chosen \emph{RAND} could contain a `correct' \emph{SMAC}, triggering an unauthorised IMSI change. However, the probability of such an event is vanishingly small, and certainly orders of magnitude smaller than the probability of a USIM failure. We therefore do not consider it further here.

\section{Analysis} \label{analysis}

The above proposals raise privacy and availability issues, which we now discuss.

\subsection{User privacy} \label{user-privacy}

The use of multiple IMSIs does not provide a complete solution to user identity confidentiality. While it is in use the IMSI still functions as a pseudonym, potentially enabling the interactions of a single phone to be tracked for a period. Of course, the more frequently IMSIs are changed the less the impact of possible tracking, but frequent IMSI changes have an overhead in terms of database management. The use of a predefined set of IMSIs further restricts the degree to which user identity confidentiality is protected. In this case, over a period of time it might be possible for an eavesdropper to link at least some of the fixed IMSIs.  \\

The design of the proposed schemes ensures that an eavesdropper will not be able to infer any confidential information from the value of \emph{RAND}\@. As discussed in sections \ref{HN_Init} and \ref{Proposal-MM-IMSI}, in schemes where the \emph{RAND} is used to signal to the USIM, the \emph{RAND} is constructed in such a way that it is indistinguishable from a truly random value; this is based on the assumption that a MAC generated $f$1 and a data string encrypted using the output from $f$5 are indistinguishable from random data. Moreover, in the scheme described in section \ref{Proposal-MM-IMSI} where the IMSI is sent embedded in \emph{RAND}\@, the IMSI (actually the MSIN) is encrypted to prevent an eavesdropper observing it.    \\

Overall the IMSI-changing proposal can be seen as allowing a trade-off between user privacy and the cost of implementing frequent IMSI changes.

\subsection{IMSI synchronisation} \label{IMSI-synch}

If, in the modifiable multiple IMSIs case, an active attacker is able to persuade the USIM to change its IMSI to an unauthorised value, then the USIM (and the UE) will cease to be able to access the network. It is therefore essential that robust cryptographic (and other) means are used to guarantee the correctness and timeliness of the new IMSI. \\

In the two variants of the predefined IMSIs scheme described in section \ref{Proposal-PM-IMSI}, loss of synchronisation cannot arise, as even if the USIM is persuaded to make an unauthorised change the new IMSI will be known to the home network.  In any event, as argued in section \ref{HN_Init}, the probability of such an event is vanishingly small. \\

Loss of synchronisation appears to be a more significant threat in the case where new IMSIs are sent embedded in the \emph{RAND} value, as in the scheme described in section \ref{Proposal-MM-IMSI}. However, as discussed there, for similar arguments to those given in section \ref{HN_Init}, the probability of a random \emph{RAND} giving a correct \emph{SMAC} is negligible. Also, malicious changes to a valid \emph{RAND}, e.g.\ involving changing the encrypted MSIN whilst leaving the \emph{SMAC} unchanged, will be detected by the AKA network authentication process.

\section{Related Work} \label{RW}

To the best of the authors' knowledge, no user privacy enhancing scheme for mobile telephony has previously been proposed that does not require changes to the existing networks. While other authors observe that significant changes to widely deployed infrastructure are unlikely to be feasible \cite{LTE-Identity-Privacy_Choudhury_2012,koien_2013_privacy}, realistic and practical proposals have not been made. Choudhury et al.\ \cite{LTE-Identity-Privacy_Choudhury_2012} have proposed a scheme to improve user identity confidentiality in the LTE network. Their scheme involves significant changes to the air interface protocol. They propose the use of a frequently changing dynamic mobile subscriber identity (DMSI) instead of the IMSI across the air interface. The DMSIs are managed by the home network and the USIM. However, the use of the DMSI imposes changes in the protocol messages, mobile equipment, and the serving network. K{\o}ien \cite{koien_2013_privacy} has recently proposed a privacy enhanced mutual authentication scheme for LTE. Although the author claims to use existing signalling mechanisms, the author introduces identity based encryption to encrypt the IMSI when sent across the air interface. \\

A number of schemes involving the use of multiple IMSIs with a single USIM have been proposed. Marsden and Marshall \cite{patent_multi-IMSI} describe a scheme to use multiple IMSIs for multiple networks with a single USIM. Their scheme involves the use of an update server to provide a suitable IMSI as and when it is required. Tagg and Campbell \cite{patent_identity} propose a similar approach involving multiple virtual USIMs on a single device. In both cases the objective is to avoid roaming charges by dynamically switching network provider. The focus of their work is thus very different to the schemes described above; they also do not provide a means of transferring new IMSIs transparently to a USIM.\\

Dupr\'e \cite{patent_controlled_SIM} presents a process to control a subscriber identity module (SIM) for mobile phone systems. He provides generic guidance regarding the transmission of control information from the network to the SIM. The schemes described in this paper extends Dupr\'e's idea in a more concrete way. 

\section{Conclusions} \label{Conclusion}

In this paper we propose two approaches to using multiple IMSIs for a mobile telephony subscriber. The goal of these proposals is to improve user privacy by reducing the impact of IMSI disclosure on the air interface. The approaches do not require any changes to the existing deployed network infrastructures, air interface protocols or mobile devices. The overhead introduced is modest and should be feasible to manage in real-world networks. One major advantage is that the proposed schemes could be deployed immediately since they are completely transparent to the existing mobile telephony infrastructure. \\

The proposed schemes provide a form of pseudonymity on the air interface, even when it is necessary to send the IMSI in cleartext. The schemes reduce the impact of user privacy threats arising from IMSI capturing. In doing so they improve user location confidentiality, because the location update message involves sending the cleartext user identity (i.e.\ the IMSI or TMSI) across the air interface.\\

There is ample scope for future work. Evaluating the proposed schemes by simulation is the current priority. Future work could concentrate on setting reliable rules for triggering an IMSI change.

\bibliographystyle{abbrv}
\balance
\bibliography{Improving_User_Privacy}


\end{document}